\documentclass[]{article}
\usepackage{emulateapj}
\usepackage{multicol}
\usepackage{psfig}
\lefthead{CHARGED-PARTICLE MOTION}
\righthead{JONES, JOKIPII \& BARING}
\eqsecnum 
\submitted{Astrophysical Journal, in press (1998)}

\begin{document} 
\title{Charged-Particle Motion in Electromagnetic Fields \\
        Having at Least One Ignorable Spatial Coordinate}

   \author{Frank C. Jones}
   \affil{Laboratory for High Energy Astrophysics, Code 660, \\
      NASA Goddard Space Flight Center, Greenbelt, MD 20771, U.S.A.\\
      \it frank.c.jones@gsfc.nasa.gov\rm}

   \vskip 5pt

   \author{J. Randy Jokipii}
   \affil{Department of Planetary Sciences, Campus Box 440,\\
      University of Arizona, Tucson, AZ 85721\\
      \it jokipii@lpl.arizona.edu\rm}

  \and

   \author{Matthew G. Baring\altaffilmark{1}}
   \affil{Laboratory for High Energy Astrophysics, Code 661, \\
      NASA Goddard Space Flight Center, Greenbelt, MD 20771, U.S.A.\\
      \it baring@lheavx.gsfc.nasa.gov\rm}

\altaffiltext{1}{Universities Space Research Association}
\date{\today}

\begin{abstract} 
We give a rigorous derivation of a theorem showing that charged
particles in an arbitrary electromagnetic field with at least one
ignorable spatial coordinate remain forever tied to a given
magnetic-field line. Such a situation contrasts the significant
motions  normal to the magnetic field that are expected in most real
three-dimensional systems. It is pointed out that, while the
significance of the theorem has not been widely appreciated, it has
important consequences for a number of problems and is of particular
relevance for the acceleration of cosmic rays by shocks.
\end{abstract}


\section{Introduction}
   \label{introduction}

The problem of particle motion in electromagnetic fields is fundamental
and  pervades both laboratory and astrophysical plasma physics.  In
spite of the conceptual simplicity of the problem, theoretical analysis
and numerical modeling of the particle motions can be difficult.
Often, in order either to simplify the analytical work or to reduce the
demand on computer resources, the system of interest is considered in
only one or two spatial dimensions, with the hope that no essential
physics will be lost.  It is readily shown that this approximation does
have a number of important general consequences which can result in the
loss of essential physics, and it must therefore be used with extreme
caution.  On the other hand, some real physical systems have a symmetry
such that one (or even two) coordinates are ignorable. In such cases
this imposes a real constraint on the possible motions of charged
particles.  While we will discuss an example of this type of real
symmetry, the primary emphasis of this paper will be to address the
profound effect  that this reduction of dimensionality has on
theoretical studies of particle acceleration.

Jokipii, K\'ota and Giacalone (\cite{jkg}), hereinafter called JKG,
presented a general theorem regarding spatial constraints to
charged-particle motion in an electromagnetic field which has at least
one ignorable spatial coordinate, and discussed explicitly the
application of this theorem to hybrid simulations of collisionless
shock waves.  In the context of hybrid plasma simulations the result
has been  stated in previous papers (\cite{thom88,thom91}), so it has
been known for a while by some workers in the field. In auroral and
radiation belt physics it has been known and applied for much longer
(\cite{storm}). The essence of this theorem is that in such systems a
charged particle is effectively forever tied, in a sense which we will
define later, to the same magnetic line of force, except for  motion
along the ignorable coordinate, or for the case of a magnetic field
entirely aligned  in the ignorable direction.  Cross-field motion is
expected in many real, three-dimensional systems (\cite{giac}).  This
constraint on particle motion represents a critical loss of physics in
many situations in which the ignorability of the coordinate is an
artifact of the analysis rather than a property of the real system.
Since JKG presented the theorem in a heuristic manner, and, because it
appears that its implication has not been always  understood or
believed, in the present paper we present a rigorous derivation of the
theorem and include an expanded discussion of its implications.

The most salient application of the theorem concerns one- and
two-dimensional hybrid and full plasma simulations, which have been
employed by numerous researchers to study particle acceleration in
collisionless shocks.  A general conclusion of these investigations is
that while the simulations can generate substantial non-thermal
populations of particles in quasi-parallel shock systems, none have
been able to demonstrate any significant acceleration of a stochastic
nature of ions and electrons in quasi-perpendicular shock
environments.  Hybrid simulations that model the quasi-parallel portion
of the Earth's bow shock (Trattner \& Scholer 1991, Scholer, Trattner
\& Kucharek 1992) produced nice comparisons with AMPTE suprathermal ion
observations, following hard on the heels of the successful modeling of
this data by Ellison, M\"obius and Paschmann (1990) using a Monte Carlo
technique to describe particle convection and diffusion.   In contrast,
highly oblique and quasi-perpendicular hybrid simulation models of the
acceleration of interstellar pick up ions at the heliospheric
termination shock (e.g.  Liewer, Goldstein and Omidi 1993, Kucharek and
Scholer 1995, Liewer, Rath and Goldstein 1995) fail to create any
measurable non-thermal ions for field obliquities above around
$50^\circ$.  For full plasma simulations, which have had only very
limited application to shock acceleration problems, the situation is
similar.  It has been concluded by such studies (\cite{Galetal}) that
perpendicular relativistic electron-positron shocks, such as would
occur in pulsar wind termination shocks, are incapable of accelerating
the ambient particles to non-thermal energies.  This situation is
modified slightly by the inclusion of a proton component (Hoshino et
al. 1992) that spawns magnetosonic waves that can provide limited
(second order stochastic) acceleration of positrons only.  We suggest
that, in light of the following discussion, such results may be viewed
as the consequence of the reduced dimensionality and the concomitant
loss of cross field diffusion.

\section{The theorem, its specification and meaning}
   \label{theorem}

The theorem states that for a system of electromagnetic fields that
have at least one ignorable coordinate the corresponding component of
canonical momentum is conserved. The conservation of this component can
have serious implications for the allowed motions of particles in such
a field system {\it e.g.} binding a particle to a given {\em field line
equivalence class} (FLEC) which is defined to be the class of all
magnetic field lines that differ only by translation along the
ignorable coordinate (or a rotation if the ignorable coordinate is an
angle variable).

In the Appendix we  show that if the electromagnetic {\em fields} do
not depend on at least one coordinate, say $x_2$, in some orthogonal
(not necessarily Cartesian)  coordinate system, then a particular gauge
can be found in which the vector and scalar potentials, $\vec A$ and
$\phi$, also do not depend on this coordinate. Furthermore, insofar as
the component $A_2$ is concerned, this gauge is unique up to an overall
constant in space and time.  This implies that the Hamiltonian of a
particle in such a field will have $x_2$ as an ignorable coordinate.

We will also show in Section \ref{equivalence} that the scaled
component of the vector potential, $h_2A_2$ (a scale factor, $h_i$,
converts a generalized orthogonal coordinate, $x_i$, into conventional
length units such as centimeters) in the ignorable direction is a
constant along a magnetic field line (as well as along the $x_2$
coordinate) and thus may be used as a label for a given FLEC where a
field line equivalence class is defined as all field lines that differ
only in their $x_2$ coordinate {\it i.e.} a surface generated by moving
a field line through all values of the coordinate $x_2$. We will also
show that a physically meaningful magnetic field line motion may be
defined. This motion reduces to the usual convection of field lines for
ideal hydromagnetic systems and can not be due to a simple gauge
relabeling since no remaining gauge transformations that preserve the
symmetry can change the value of $h_2A_2$. We will also show in Section
\ref{binding} that the physical scale of this binding is of the order
of the harmonic mean of the particle's gyroradius in this field region.

\subsection{The equivalence class and its motion}
   \label{equivalence}

Since we have shown that with our choice of gauge nothing depends on
the $x_2$ coordinate we may establish an equivalence class of all field
lines that differ only in their $x_2$ coordinate.  Further, from the
definition of $\vec B$, and the fact that $\vec B \times \vec B = \vec
B \times (\vec \nabla \times \vec A) = 0$, we may take the $2$
component of $\vec B \times \vec B$ and obtain
\begin{equation}
B_1{1\over h_1}{\partial (h_2A_2) \over \partial x_1}
 + B_3{1\over h_3}{\partial (h_2A_2) \over \partial x_3}
 = \vec B \cdot \vec \nabla (h_2A_2) = 0
\end{equation}
and $h_2A_2$ is constant along a magnetic field line.  This means that
we may {\em identify} a field line (or rather its equivalence class) by
its value of $h_2A_2$. Furthermore, if, at a later time, the FLEC with
this value of $h_2A_2$ is found at a different spatial position we may
{\em define} this motion as a field line motion knowing that this
motion is uniquely determined by the physics because there is no
remaining time dependent gauge freedom that  changes the value of
$A_2$.

Employing the electric and magnetic fields we define the velocity
\begin{eqnarray}
\vec w &=& {c\over B_1^2+B_3^2} E_2(\hat{e}_1B_3 - \hat{e}_3B_1)
     \nonumber\\[-5.5pt]
 && \\[-5.5pt]
    &=& -{\partial (h_2A_2) \over \partial t}{\vec \nabla (h_2A_2) \over 
        (\vec \nabla (h_2A_2))^2}\; ,\nonumber
\end{eqnarray}
where $\hat{e}_1$ and $\hat{e}_3$ are unit vectors in the $x_1$ and $x_3$ 
directions, respectively, it is easy to see that $\vec w$ satisfies
\begin{equation}
{\partial (h_2A_2) \over \partial t} + \vec w \cdot \vec \nabla (h_2A_2) = 0
\end{equation}
and may, therefore, be regarded as the velocity, perpendicular to the
$x_2$ direction, of the field lines equivalence class, defined by their
value of $h_2A_2$. It is also clear that this velocity is identical to
the $\vec E \times \vec B$ drift velocity whenever $\vec E \cdot \vec B
=0$.

\subsection{Binding the particle to a field line (Equivalence class)}
   \label{binding}

From this point the argument is nearly the same as JKG, and it is
straightforward to see that a particle is  ``bound'' in some sense to a
particular value of $h_2A_2$ and hence to a field line.  Since $x_2$ is
an ignorable coordinate the $x_2$ component of the canonical momentum
is conserved (\cite{gold}), $h_2p_2 + {e \over c} h_2A_2 = {\rm const}$.
Therefore, since $-p \leq p_2 \leq p$, we have
\begin{equation}
   (h_2A_2)_{\rm Max} - (h_2A_2)_{\rm Min} \leq {2cp\over e} \bar{h}_2\; ,
\end{equation}
where $\bar{h}_2 \equiv [(h_2)_{\rm Max}+  (h_2)_{\rm Min}]/2$.  In
situations where the particles energy and hence $p \ \ (\ \ =
\sqrt{h_1^2p_1^2+h_2^2p_1^2+h_3^2p_3^2}\ \ )$ do not change this
defines a flux tube, in which the particle is constrained to remain.

To obtain an estimate of the spatial scale of this flux tube consider a
curve, $S$, in the $x_1,x_3$ plane, from the field line with the
minimum value of $h_2A_2$ to the one with the maximum value, that is
everywhere perpendicular to the field $\vec B^\prime(x_1,x_3) \equiv
\hat{e}_1 B_1(x_1,x_3) +\hat{e}_3 B_3(x_1,x_3)$ where $d\vec s \cdot
\vec B^\prime = 0$ along the curve $S$. We have then
\begin{eqnarray}
  && (h_2A_2)_{\rm Max} - (h_2A_2)_{\rm Min} \nonumber\\[3pt]
  &&\qquad =\; \int_Sdx_1~{\partial (h_2A_2)\over\partial x_1} 
         +\int_Sdx_3~{\partial (h_2A_2)\over\partial x_3} \\[3pt]
  &&\qquad =\; -\int_S (h_2d\vec s \times \vec B^\prime)_2
           \;\geq\; (h_2)_{\rm Min}\int_S ds~|B^\prime|\nonumber
 \label{integral1}
\end{eqnarray} 
employing Equation (\ref{curve_curl}).  Since the absolute value of
this integral $\leq{2cp \over e} \bar{h}_2$ we may divide equation
(\ref{integral1}) by this quantity to obtain
\begin{equation}
\int_S {ds\over r_g^\prime} \leq \left( 1+ {(h_2)_{\rm Max}
 \over (h_2)_{\rm Min}}\right)
 \label{integral2}
\end{equation}
where $r_g^\prime \equiv {cp / eB^\prime}$ is the local gyroradius of a
particle due to the $x_1$ and $x_3$ components of the magnetic field.
A particle, therefore, moves a distance perpendicular to the $x_1,x_3$
component of the order of the harmonic mean of its gyroradius in this
field along this path. Clearly, the electromagnetic fields may be
arbitrary, consistent with Maxwell's equations and reduced
dimensionality.

It should be noted, however, that if $\vec B$ lies entirely in the
$x_2$ direction over a region of space, $\partial (h_2A_2) / \partial
x_3$ and $\partial (h_2A_2) / \partial x_1$ are zero and $r_g^\prime
\longrightarrow \infty$.  There is, therefore, no restriction,  imposed
by the theorem, on particle motion in this region.

\section{Specific consequences}
 \label{consequences}

The theorem established in the previous section makes it clear that a
full discussion of any problem involving charged particles and
electromagnetic fields must either be done in three dimensions or the
effects of reduced dimensionality must be shown to be irrelevant to the
problem at hand.  This has important consequences for a number of
problems, including the physics of collisionless plasmas and cosmic-ray
transport.  We consider briefly some specific applications involving
both physically real and artificially constrained symmetries.

\subsection{L Shell Conservation}

We will begin by considering a case in which an identifiable symmetry
reflects (at least approximately) a {\em real} property of the system.
It has been known for many years that the particles trapped in the
geomagnetic field (the Van Allen belts) drift around the earth in such
a way as to approximately conserve the ``L'' parameter. This parameter
is essentially a label for the FLEC generated by the rotational
symmetry of the dipole field. Since this symmetry is broken in many
ways  (the dipole is tilted with respect to the earth's rotation axis
and many perturbations of the field, both time dependant and static, do
not preserve this symmetry) the conservation is only approximate but is
good enough to be a useful concept in space physics. Furthermore the
entire concept of trapped (radiation belts) and excluded particles
(cosmic-ray geomagnetic cutoff) was discussed by St\"{o}rmer
(\cite{storm}) in terms of the ``second first integral'' of the motion,
which is clearly the azimuthal component of the canonical momentum.  In
addition, the motion of the particles in the remaining degrees of
freedom can often be understood by viewing the conserved momentum in
the Hamiltonian as part of a potential function (Stern 1975).  In such
a case, the actual symmetry of the system aids in the understanding of
particle dynamics.

\subsection{Problems in Charged Particle Transport}
    \label{transport}

Previous discussions of cosmic-ray transport using the quasi-linear
approximation (\cite{jok66}) often assumed irregularities which
depended only on one coordinate (slab turbulence).  In such geometries
the field-line mixing, or random walk was the {\it only} contribution
to motion normal to the average magnetic field.  Yet, other studies of
phenomenological particle scattering in a magnetic field
(\cite{fj&o,jon90}) showed that cross field diffusion should occur. It
is now clear that the reduced dimensionality of the slab model, not the
quasi-linear approximation, guaranteed this result and  since real
turbulence is usually three dimensional in nature (\cite{bw&m}) this
limitation is probably not real, although the field line mixing
probably plays an important role.

Recently Giacalone and Jokipii (1994) considered a number of issues in
the general problem of cosmic-ray transport and its relation to the
dimensionality of magnetic-field fluctuations.  Using synthesized
magnetic-field turbulence having a specific power spectrum, they
directly verified that particles cannot cross field lines if the
spectrum does not depend on all three spatial variables.  However, when
the turbulence was allowed to be  fully three dimensional, cross field
diffusion of the particles was evident.  This has profound implications
for simulations of particle acceleration by collisionless shocks.

\subsection{Collisionless Shocks}
   \label{shocks}

As discussed in the introduction, both hybrid and full particle one-
and two-dimensional simulations do not predict any populations of ions
accelerated to highly suprathermal energies in quasi-perpendicular
shocks,  This provides a major problem for these approaches: they are
in direct conflict with ubiquitous observations of such populations
near highly oblique traveling interplanetary shocks (e.g. Baring et al.
1997), and the general consensus that the anomalous cosmic ray
populations result from the acceleration of pick-up ions at the solar
wind termination shock (e.g. Pesses, Jokipii \& Eichler 1981).  The
discussion of JKG pointed out that the important problem of
charged-particle acceleration at quasi-perpendicular collisionless
shocks could not be properly studied using one- or two-dimensional
hybrid simulations.  This is because cross-field motion plays a crucial
role.  The restrictions for the acceleration at a purely perpendicular
shock are particular severe.  Here the magnetic field is carried
downstream at the post-shock flow speed.  If the particles are to
remain tied to a magnetic field line, as required by the theorem, any
acceleration at the shock would have to be fast enough for the particle
gyro-radius to increase as fast as the distance of the magnetic field
line downstream.  Otherwise the particle would have to be convected
downstream with the fluid, preventing any further acceleration.  It is
readily shown that the ordinary gradient drift along the shock face is
too slow to effect the required increase in the gyroradius; however,
see the discussion in Section \ref{surfing}.  Hence significant
particle acceleration depends critically on the charged particles being
able to scatter and diffuse across the magnetic field, so that they may
continue interacting with the shock.  This necessary motion is
artificially suppressed in simulations of reduced dimensions.
Therefore one would incorrectly infer that significant acceleration
cannot occur in such quasi-perpendicular shocks.  The situation in
oblique shocks is not so severe since particles may be able to follow
field lines to continue interacting with the shock.  But in such cases
the quantitative character of the results will be affected because the
cross-field transport is still important for efficient injection.

Furthermore, simulations of {\it quasi-parallel} shocks, where it has
been noted that acceleration occurs by drifting along a transverse
fluctuation in the magnetic field (\cite{kands}), should also be
re-examined using three-dimensional simulations.  In fact, Kucharek,
Fujimoto \& Scholer (1993) demonstrated that significant changes are
incurred in quasi-parallel shock simulations in going from
one-dimensional to two-dimensional constructions.

\subsection{Acceleration by Field Line Merging and Shock Surfing}    
   \label{surfing}

As examples of situations in which the conservation of the canonical
momentum component does {\em not} prevent particle acceleration we
examine two cases where this conservation can actually {\em cause} the
acceleration.  There exist situations in which a charged particle is
prevented from remaining near to the FLEC on which it was once found.
This can arise because the FLEC ceases to exist (field line merging) or
a strong electric field temporarily prevents the particle from
following the field line motion (shock surfing).

For the first example consider a planar current sheet in the $y-z$
plane which separates two regions of oppositely-directed magnetic
field, $\vec B=\vec e_z B_0 {\rm sgn}(x)$, where $B_0$ is a constant.
We can choose $A_y(x) = B_0 \vert x \vert$ so that field lines that are
equidistant from the neutral sheet but on opposite sides (therefore of
opposite sign) will have the same value of $A_y$.  A given particle in
the vicinity of this neutral sheet could range a further distance in
$x$ than it could if the field did not change sign but must still
maintain the constancy of the quantity $p_y + {q\over c} A_y$.

If now, as shown in Figure \ref{merge},  the regions of oppositely
directed fields are allowed to merge and annihilate field through
plasma processes, which we do not discuss here, the value of $A_y$ will
everywhere continuously increase due to the constant loss of the
smallest values in the vicinity of $x=0$.  In order to preserve the
invariant value of $p_y + {q\over c} A_y$, the value of $\vert
p_y\vert$ must continuously increase.  Thus $p$ must continuously
increase and the particle must be  accelerated by the annihilating
fields.  Ambrosiano, et al. (\cite{ambro}) employ an argument similar
to the above in a discussion of particle acceleration.

\vskip 1.0truecm
\centerline{}


\centerline{}
\vskip -0.2truecm
\centerline{\psfig{figure=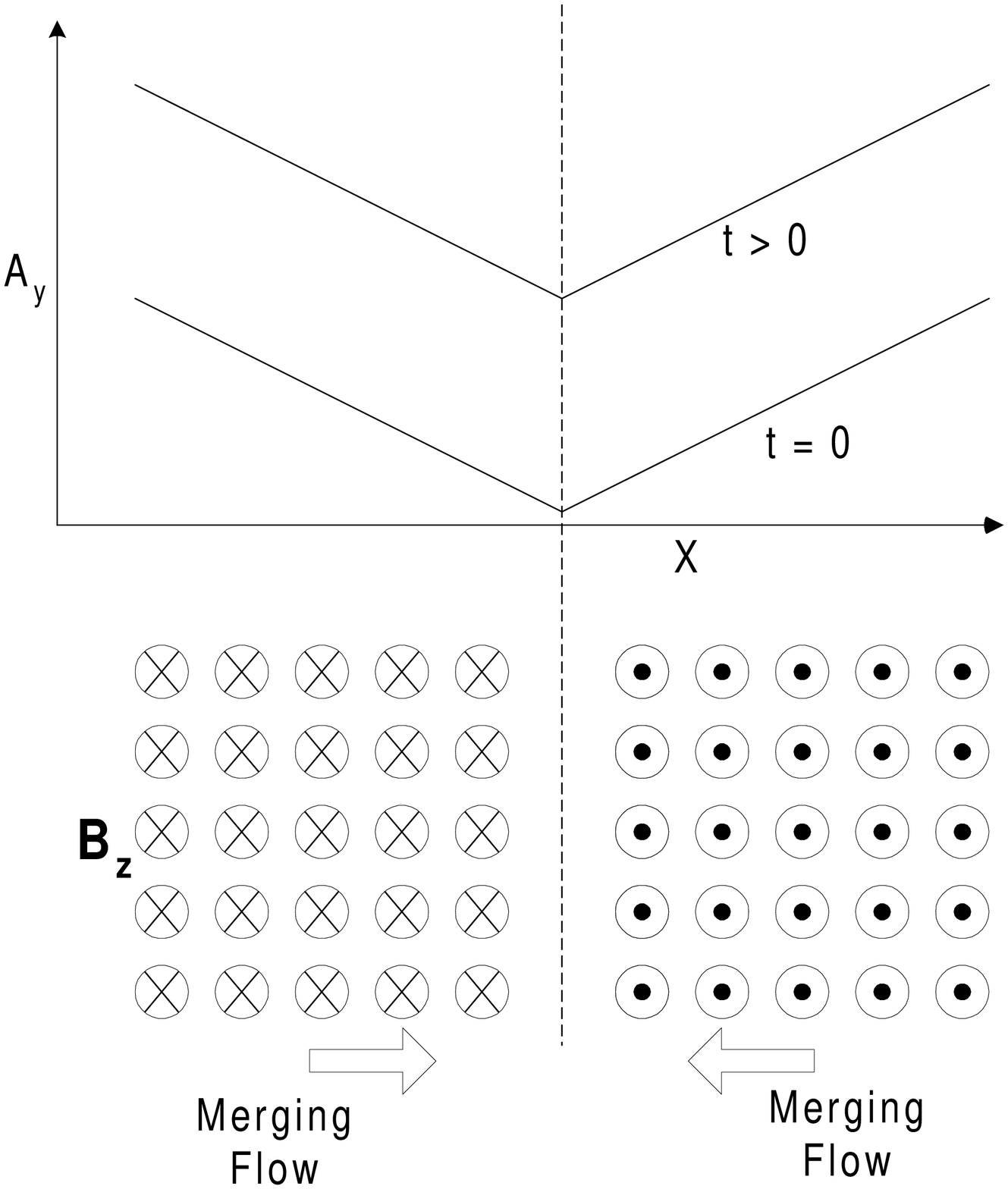,width=8.0cm}}
\vskip -0.6truecm
   \figcaption{Motion of field lines in a region of field line merging   
     (annihilation) showing the increase of the local value of $A_y$.
\label{merge}}
\centerline{}

In a study of plasma shocks with a cross shock electric potential drop
it has been shown (\cite{zank_etal,lee_etal}) that some particles that
are incident on the shock front will not have enough kinetic energy to
overcome this potential and thus can not cross the shock  because the
sign of the potential drop is always in the direction to retard the
incoming ions (\cite{GandS,JandE87}).  Once again, if there is an
ignorable direction perpendicular to the magnetic field the particle
must remain within a gyroradius or so of it's original FLEC, which is
proceeding through the shock front without the particle. The only way
this can occur is for the particle's gyroradius to increase as fast as
the field lines move downstream. This process, called shock surfing
because the particle skims along the shock surface along the ignorable
direction, continues until the particle gains sufficient energy to
overcome the cross shock potential.

Since both of these mechanisms depend rather strongly on there being no
variation of the electric and magnetic fields in the ignorable
direction any perturbation that does not have that symmetry can
``derail'' the process and limit the energy gain to one considerably
less than the theoretical limit based on that symmetry.  Therefore,
once again, theoretical investigations of these processes should allow
for variations in three dimensions to determine the true limits on
their effectiveness as accelerating mechanisms.

\section {Conclusions}

In this paper we have provided a rigorous proof of the JKG theorem and
emphasized that the most important application of the theorem is that
the assumption of reduced dimensionality often used in modeling
charged-particle motion has severe effects on the validity of the
conclusions which may be drawn.  In particular, any processes which may
depend on particle motion across the magnetic field cannot in principle
occur.  This therefore casts doubt on those models or simulations which
are complex enough that the contribution of cross-field motion cannot
be independently assessed.  In such cases the only recourse is to do a
fully three-dimensional model.

\section*{Acknowledgments.}

We are grateful to D. Ellison, J. Giacalone, C. Graziani and J. K\'ota
for valuable discussions.  This work was supported by the NASA Space
Physics Theory Program under grant NAGW 50552251.  JRJ was also
supported in part by NASA under grant NAGW 1931 and by NSF under Grant
ATM 559307046.

\appendix
\section{Proof that a suitable gauge exists}

Since there exist many symmetries other than linear translation for which
electromagnetic fields are invariant we consider an arbitrary,
curvilinear, orthogonal coordinate system with coordinates $x_1$, $x_2$
and $x_3$. In general such a coordinate system the line element is
given by
\begin{equation}
\vec{ds} = \hat{e}_1h_1dx_1 +\hat{e}_2 h_2dx_2 + \hat{e}_3 h_3dx_3
\end{equation} 
where the $\hat{e}_i$ symbol is the  unit vector in the $i$th direction
and $h_1,h_2$, and $h_3$ are the length scale factors for the three
coordinates (for a Cartesian coordinate system $h_1=h_2=h_3=1$.) In
this system we have
\begin{eqnarray}
\vec{B} &=& \vec \nabla \times \vec{A} 
= {1\over h_1h_2h_3}\left[h_1\hat{e}_1\left({\partial (h_3A_3) \over \partial x_2} - {\partial (h_2A_2) \over \partial x_3} \right)\right. \nonumber \\ 
&-& \left. h_2\hat{e}_2\left({\partial (h_3A_3) \over \partial x_1} - {\partial (h_1A_1) \over \partial x_3} \right) + h_3\hat{e}_3\left({\partial (h_2A_2) \over \partial x_1} - {\partial (h_1A_1) \over \partial x_2} \right) \right]  
 \label{curve_curl}
\end{eqnarray}
Consider an electromagnetic field $\vec E(\vec r,t),\vec B(\vec r,t)$
in which a particle of charge $e$ and rest mass $m_0$ moves with
velocity $\vec v$.  The only restriction is that $\vec E$ and $\vec B$
are independent of one of the spatial coordinates, say $x_2$.

We first demonstrate explicitly that if the electromagnetic field is
not a function of  a given coordinate the vector potential $\vec A$ may
be chosen to be independent of this coordinate as well.  The first
point to note is that this independence is manifest by the fact that
the field structure may be moved an amount $ds$ in the direction of
this ignorable coordinate (if the coordinate is an angle variable the
motion is a rotation) and the resulting field will be identical to the
original. We emphasize that the transformations that follow are {\em
not} coordinate transformations; they are translations or rotations of
the fields, potentials and gauge fields themselves. The transforms are
{\em not} applied to Maxwell's Equations, rather, Maxwell's Equations
are applied to the transformed fields. Therefore an infinitesimal
transformation gives
\begin{eqnarray}
&&{\delta\vec B\brace\delta\vec E} = ds~\hat{e}_2\cdot\vec\nabla {\vec B\brace\vec E}={ds\over h_2}{\partial\over \partial x_2}{\vec B\brace\vec E}=0\nonumber \\
&&{\delta\vec A\brace\delta\phi} = ds~\hat{e}_2\cdot\vec\nabla{\vec A\brace\phi} ={ds\over h_2}{\partial\over \partial x_2}{\vec A\brace\phi}\not= 0~~~ {\rm (in~general)}
\end{eqnarray}
where $ds=|\vec{ds}|$.  Since 
\begin{eqnarray}
\delta \vec B = \vec B^\prime - \vec B &=& \vec\nabla \times (\vec A^\prime - \vec A) \nonumber\\
&=& \vec\nabla \times (\vec{\delta A}^\prime)
\end{eqnarray}
and 
\begin {equation}
\vec A^\prime = \vec A + ds{\partial \vec A \over h_2 \partial x_2}
\end{equation}
we have
\begin{equation}
\delta \vec B=ds~ \vec \nabla \times {1\over h_2}{
\partial \vec A \over \partial x_2} = 0,
\end{equation}
so we may express the partial derivative with respect to $x_2$ as a potential field because it's curl vanishes;
\begin{equation}
{1 \over h_2 }{\partial \vec A \over \partial x_2} = -\vec \nabla \Psi
\end{equation}
where $\Psi$ is a scalar function.  Since, similarly, for $\vec E$ we
have
\begin{equation}
\delta\vec E= \vec E^\prime - \vec E = -\vec\nabla(\delta\phi) - {1\over c}{\partial \over \partial t}(\delta\vec A) = 0,
\end{equation}
the condition on $\vec E$ is
\begin{eqnarray}
{1\over h_2}{\partial \vec E \over \partial x_2} &=& -\vec \nabla {1\over h_2}{\partial \phi \over \partial x_2} - {1\over c} {\partial \over \partial t}{1\over h_2}{\partial \vec A \over 
\partial x_2}\\
&=& -\vec \nabla\left({1\over h_2}{\partial \phi \over \partial x_2} - {1\over c}{\partial 
\Psi \over \partial t} \right) = 0.
\end{eqnarray}
This requires 
\begin{equation}
{1\over h_2}{\partial \phi \over \partial x_2} - {1\over c}{\partial 
\Psi \over \partial t} = F(t) \label{phi cond}
\end{equation}
where $F(t)$ is a function of $t$ alone.

We now may make a gauge transformation
\begin{eqnarray}
\vec A &\rightarrow& \vec A^{\prime} = \vec A + \vec \nabla \lambda\nonumber \\ 
\phi &\rightarrow& \phi^{\prime}  = \phi - {1\over c} {\partial \lambda \over\partial t}
\end{eqnarray}
where $\lambda$ is an arbitrary scalar function.  

If we choose 
\begin{equation}
\lambda = \int_{C}^{x_2} \Psi h_2 ~dx^\prime_2 +  K(t)\int_{C}^{x_2}h_2 ~dx_2^\prime  , \label{lambda choice}
\end{equation}
where $\int_{C}^{x_2}$ is an integral running from a fixed surface with $x_2 = {\rm const.}$ along a curve of constant $x_1,x_3$ and $K(t)$ is an arbitrary function of time alone, we see that 
\begin{equation}
{1\over h_2}{\partial \vec A^{\prime} \over \partial x_2} = {1\over h_2}{\partial \vec A \over 
\partial x_2} + \vec\nabla {1\over h_2}{\partial \lambda \over \partial x_2} = 
\vec \nabla\left({1\over h_2}{\partial \lambda \over \partial x_2} - \Psi\right) = 0
\end{equation}
and we have gauged away any $x_2$ variation of $\vec A^{\prime}$.

Similarly we have using equations \ref{phi cond} and \ref{lambda choice}
\begin{equation}
{1\over h_2}{\partial \phi^{\prime} \over \partial x_2} = {1\over h_2}{\partial \phi \over \partial 
x_2}  - {1\over c}{\partial \over \partial t}{1\over h_2}{\partial \lambda \over
\partial x_2}  = {1\over h_2}{\partial \phi \over \partial x_2} -{1 \over c} {\partial 
\Psi \over \partial t} - {1 \over c}{ \partial K(t) \over \partial t} = 
F(t) - {1 \over c}{ \partial K(t) \over \partial t}\, ,
\end{equation}
so if we choose
\begin{equation}
K(t) = \int^t cF(t^\prime) dt^\prime + K_0
\end{equation}
where $K_0$ is an arbitrary constant, we may also gauge away the $x_2$ 
variation in $\phi$ as well.  

We can readily see that the value of $A_2$ is fixed, up to an overall
constant in space and time, by the requirement that both $\vec A$ and
$\phi$ are independent of the $x_2$ coordinate. For if one performs a
further gauge transformation
\begin{eqnarray}
\vec A & \rightarrow & \vec A^{ \prime} = \vec A + 
\vec \nabla \Lambda\\
\phi & \rightarrow & \phi^{ \prime} = \phi -
{1\over c}{\partial \Lambda \over \partial t}
\end{eqnarray}
while requiring
\begin{eqnarray}
{1\over h_2}{\partial \vec A \over \partial x_2} &=& {1\over h_2}{\partial \vec A^{ 
\prime} \over \partial x_2} = 0~~{\rm and}\nonumber \\[3pt]
{1\over h_2}{\partial \phi \over \partial x_2} &=& {1\over h_2}{\partial \phi^{ \prime} \over \partial x_2} = 0
\end{eqnarray}
we immediately obtain the requirement
\begin{eqnarray}
\vec \nabla\left({1\over h_2}{\partial \Lambda \over \partial x_2}\right) &=& 0 \nonumber \\[3pt]
{1\over c} {\partial \over \partial t}\left({1\over h_2}{\partial \Lambda \over \partial x_2}\right) &=& 0
\end{eqnarray}
and $h_2^{-1}{\partial \Lambda / \partial x_2}$  ($= A^\prime_2 - A_2$) is a constant in space and time, which is merely a new choice of $K_0$.

\end {document}